# Thermal transport in suspended silicon membranes measured by laser-induced transient gratings

A. Vega-Flick [1,2], R. A. Duncan [1], J. K. Eliason [1], J. Cuffe [3], J. A. Johnson [1,4], J.-P. M. Peraud [3], L. Zeng [3], Z. Lu [3], A. A. Maznev [1], E. N. Wang [3], J. J. Alvarado-Gil [2], M. Sledzinska [5], C. M. Sotomayor Torres [5,6], G. Chen [3], K. A. Nelson [1].

[1] Department of Chemistry, Massachusetts Institute of Technology, 77 Massachusetts Avenue, Cambridge, Massachusetts 02139, USA.

[2] Applied Physics Department, CINVESTAV-Unidad Mérida, Carretera Antigua a Progreso Km 6, Cordemex, Mérida, Yucatán 97310 Mexico.

[3] Department of Mechanical Engineering, Massachusetts Institute of Technology, 77 Massachusetts Avenue, Cambridge, Massachusetts, 02139, USA.

[4] Department of Chemistry and Biochemistry, Brigham Young University, Provo, Utah 84602, USA.

[5] Catalan Institute of Nanoscience and Nanotechnology (ICN2), CSIC and The Barcelona Institute of Science and Technology, Campus UAB, Bellaterra, 08193 Barcelona, Spain.

[6] ICREA, Pg. Lluís Companys 23, 08010 Barcelona, Spain.

Studying thermal transport at the nanoscale poses formidable experimental challenges due both to the physics of the measurement process and to the issues of accuracy and reproducibility. The laser-induced transient thermal grating (TTG) technique permits non-contact measurements on nanostructured samples without a need for metal heaters or any other extraneous structures, offering the advantage of inherently high absolute accuracy. We present a review of recent studies of thermal transport in nanoscale silicon membranes using the TTG technique. An overview of the methodology, including an analysis of measurements errors, is followed by a discussion of new findings obtained from measurements on both "solid" and nanopatterned membranes. The most important results have been a direct observation of non-diffusive phonon-mediated transport at room temperature and measurements of thickness-dependent thermal conductivity of suspended membranes across a wide thickness range, showing good agreement with first-principles-based theory assuming diffuse scattering at the boundaries. Measurements on a membrane with a periodic pattern of nanosized holes (135nm) indicated fully diffusive transport and yielded thermal diffusivity values in agreement with Monte Carlo simulations. Based on the results obtained to-date, we conclude that room-temperature thermal transport in membrane-based silicon nanostructures is now reasonably well understood.

## I. INTRODUCTION

Thermal transport in nanostructures and nanostructured materials has recently become an area of much interest due to novel phenomena that emerge for many materials on the nanoscale, as well as the associated implications for practical engineering efforts in the fields of thermoelectrics and microelectronics [1,2,3,4]. Much recent experimental effort has been concentrated on studying phonon-mediated thermal transport in silicon nanostructures [5,6,7,8,9,10,11,12,13,14]. The focus on silicon is explained on one hand by the practical importance of this material for many applications, and on the other hand by the fact that silicon serves as a convenient "model material" as it has been very thoroughly studied and is readily amenable to nanofabrication. Indeed, starting from early work on the phonon size effect in silicon thin films [15], experiments on Si nanostructures have yielded many important advances [2,6,17]. However, we are still far from the complete understanding of heat transport in Si nanostructures, even at room temperature. In fact, a number of issues related to recent experimental observations, such as "below Casimir limit" thermal conductivity of Si nanowires [5] and the role of "phononic crystal" effects in the thermal transport in nanoporous Si membranes [6,7] remain hotly debated.

Characterizing thermal transport in nanostructures presents a formidable challenge for metrology. Oftentimes key conclusions are drawn on the basis of absolute values of the measured thermal conductivity [5,6]. However, accurate measurements of thermal conductivity are difficult even in bulk materials, as illustrated by the effort involved in quantifying the isotope effect on the room temperature thermal conductivity of Si [18]. The difficulties are greatly amplified in thermal conductivity measurements on nanostructures: for example, the device-based approach, in which the measurement device is fabricated together with the nanostructure to be measured [19], challenges the notions of reproducibility and benchmarking in metrology. Another challenge often lies in the physics of the measurement, which typically involves metal heaters [20]. The effect of the metal heater and the associated thermal boundary resistance can be easily accounted for within the framework of the thermal diffusion model [21]. However, when the spatial dimensions become comparable to the phonon mean free path (MFP) and the diffusion model is not longer valid, the task becomes much more difficult [22,23].

The above challenges create a need for measurement techniques which, on one hand would possess inherently high absolute accuracy, and on the other hand would permit measurements of thermal transport without metal heaters or any other extraneous structures. One such technique has been known for some time under the name laser-induced transient thermal gratings [24,25]. In this method, two short laser pulses are crossed in the sample, creating a spatially periodic temperature profile or thermal grating (TTG). The decay of the TTG via thermal diffusion is monitored via diffraction of a probe laser beam. The measurement is based on the dynamics of the TTG decay and does not require knowledge of the absolute temperature rise and heat flux. The only parameter pertaining to the measurement setup is the TTG period, which can be determined with high accuracy. The measurement is entirely noncontact and nondestructive, with the sinusoidal temperature profile yielding an additional advantage by making the measurement most amenable to theoretical analysis. A number of practical aspects has been addressed by the introduction of optical heterodyne detection [26,27], which has resulted in an efficient and compact set-up.

In this paper, we review recent measurements of thermal transport in nanoscale Si membranes using the TTG technique. We begin with a discussion of the methodology and error analysis, with most attention paid to quantifying the "overheating" effect caused by the pump and probe lasers. We then review new findings obtained from measurements on both "solid" and nanopatterned membranes, relying both on published results [13,14] and recently obtained data. We conclude with a discussion of future prospects, challenges, and opportunities for studying thermal transport in nanostructures using the TTG method.

## II. METHODOLOGY

### A. Transient thermal grating technique and experimental setup

In the TTG technique, illustrated schematically by Fig. 1(a), two crossed laser pulses create an interference pattern with period $L = \lambda/2\sin(\theta/2)$ defined by the angle $\theta$ between the beams. Subsequent absorption of the laser light by the sample creates a TTG, i.e., a sinusoidal temperature profile. In the linear response regime (which holds when the temperature variation is small compared to the average background temperature) the TTG profile remains sinusoidal while its amplitude decays as thermal energy is redistributed from peaks to nulls. The TTG decay is monitored via diffraction of a probe laser beam. The heat transfer distance is controlled by the grating period $L$, which, in our setup, was varied between 1-30 $\mu$m.

In TTG measurements presented in this work, thermal transport was nearly one-dimensional, as the TTG period was always much larger than the membrane thickness (and the optical penetration depth 0.7 $\mu$m at the excitation wavelength 515 nm was also larger than the membrane thickness for most samples). For a one-dimensional TTG, the heat diffusion equation yields an exponential thermal decay [24]

$$T(x,t) = T_o(1 - \cos(qx))\exp(-t/\tau), \tag{1}$$

with the decay time

$$\tau = 1/(\alpha q^2), \tag{2}$$

were $q = 2\pi/L$ is the grating wavevector magnitude and $\alpha$ is the thermal diffusivity, related to the thermal conductivity $\kappa$ by $\alpha = \kappa/C_V$ where $C_V$ is the volumetric heat capacity. Thus the thermal diffusivity can be determined from the TTG decay time; the only other quantity needed for the measurements is the TTG wavevector, which can be determined with high accuracy.

The TTG setup used in the experiments described in this paper employs optical heterodyne detection [26,27] in which the diffracted signal is superposed with a reference beam (i.e. "local oscillator") derived from the same source as the probe. The heterodyne detection not only increases the signal to noise ratio (S/N) but also yields a signal linear with respect to the material response, simplifying the analysis and interpretation of the measurements [27,28]. Figure 1(b) shows the optical setup used in the measurements on Si membranes. A phase mask, optimized to maximize the 1st diffraction orders, is used to produce excitation and probe/reference beam pairs. A two-lenses confocal imaging system is employed to recombine the beams at the sample. In the case of 2:1 demagnification used in the setup the TTG period equals a quarter of the phase mask period. The TTG formed in the sample modulates the refractive index and caused surface displacement via thermal expansion. Both refractive index variations and surface displacement contribute to diffraction of the probe beam. The diffracted probe is overlapped with the reference beam for heterodyne detection. The phase difference between the diffracted probe and the reference beam is controlled by tilting a high-parallelism glass plate in the probe beam path. The reference beam was attenuated with a neutral density filter in order to avoid saturation of the detector. The excitation pulses have a wavelength of 515 nm wavelength, pulse duration of 60 ps and are produced at a repetition rate of 1 kHz. The 532 nm probe beam is chopped by an electro-optic modulator to produce rectangular pulses of 68 $\mu$s in order to reduce sample heating. The signal is recorded by a fast photodiode (Hamamatsu C5658, 1 GHz bandwidth) whose output is fed to an oscilloscope.

Fig. 1(c) shows typical signal traces obtained from a 340 nm thick nanoporous Si membrane (see Sec. V) at TTG periods ranging from 3.2 $\mu$m to 10 $\mu$m. The inset shows a complete time trace with a sharp negative peak due to electronic excitation. Since the ambipolar carrier diffusion coefficient for Si is an order of magnitude larger than the thermal diffusivity [29], the charge carrier dynamics are temporally separated from the thermal transport: the grating of carrier concentration is washed away much faster than the thermal grating. Typically we analyze the signal waveform using a bi-exponential fit where one decay time corresponds to the carrier dynamics and the second to the thermal transport. For the TTG periods used in this work the charge carrier decay times are typically below 3 ns while the thermal decay times range from tens to hundreds of nanoseconds; alternatively, the tail of the signal can be analyzed using a single-exponential fit, yielding nearly identical results; however, the latter approach has been found to yield a larger statistical error.

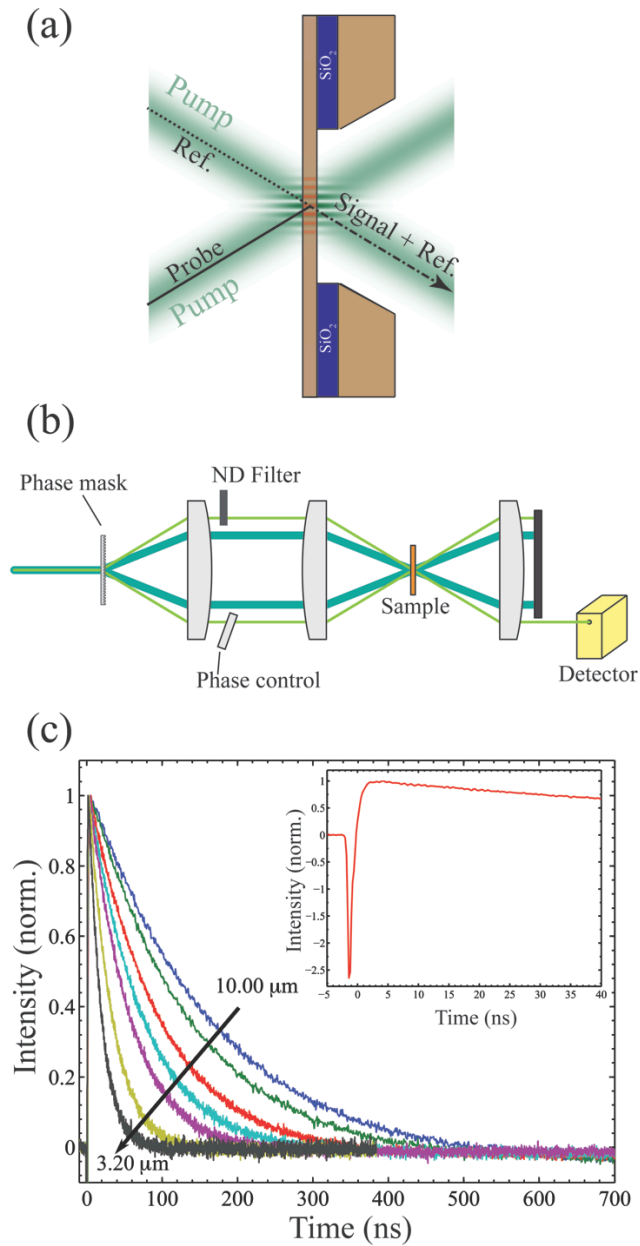

FIG. 1: (a) Transient thermal grating measurement: crossed pump beams generate a transient temperature grating, whose dynamics are measured by diffraction of a probe beam. The probe is superimposed with a reference to achieve heterodyned detection. (b) Schematic of the experimental set-up. (c) Typical signal waveforms for a range of grating periods. Inset: short-time dynamics showing a negative peak due to photoexcited carriers for a TTG period of 7.5 $\mu$m.

### B. Measurement accuracy

According to Eq. 2, the only parameter needed to determine the thermal diffusivity from the measured decay time is the TTG wavevector $q$; the knowledge of neither the magnitude of the temperature variation nor of the heat flux is required, which gives the TTG technique an advantage of inherently high absolute accuracy.

The value of $q$ is set by the period of the phase mask and the magnification of the imaging system. Due to manufacturing tolerances of the lenses the magnification is typically slightly deviated from its nominal value, and a calibration is required to ensure high absolute accuracy of the wavevector values. The calibration can be done by increasing the excitation energy above the damage threshold and burning a permanent grating pattern which is measured with a microscope (which needs, in turn, be calibrated). Alternatively, the TTG experimental setup can be used to measure the speed of sound of acoustic waves in a well-characterized sample. We have measured the surface acoustic wave (SAW) frequency along the ⟨100⟩ direction on the (001) surface of silicon and used the known SAW velocity in this direction to determine the wavevector [30]. The SAW velocity was calculated from the literature values of the elastic constants of silicon; a survey of the literature determined that these constants are known to within an accuracy of ~0.1% and therefore the TTG wavevector value is also determined to within an error on the same order. We found that the deviation of the calibrated wavevector from the nominal value determined from the phase mask period and the nominal magnification of the imaging system amounted to 0.15%.

In addition to the systematic error associated with the wavevector calibration, the accuracy of the TTG technique is affected by the error in measuring $\tau$. We assess the statistical error in $\tau$ by determining the standard deviation in a series of thermal diffusivity measurements performed at different TTG periods (in the diffusive limit, when the thermal diffusivity is expected to be independent of the TTG period, see Sec. III). The statistical reproducibility estimated in this fashion is typically 2-4%. The static repeatability determined by repeating the measurement without changing anything in the setup is typically significantly better, ~1%. The error associated with the detector response predominantly affects signals with either very fast or very slow time dynamics. Although the electronic and thermal responses are in general quite well temporally separated for silicon membranes, the persistence of an electronic decay could be a source of error depending on the trace time window. These sources of error can be taken into account by quantifying the variation in $\tau$ obtained with different time windows. This error was found to be ~2% for a set of Si membranes in a large thickness range [14].

A significant contribution to the error of TTG measurements in Si membranes is the effect of sample heating by pump and probe lasers. The thermal diffusivity is normally reduced with increasing background temperature. Therefore, laser heating of the sample results in a reduction of the measured thermal diffusivity compared to its true value at the nominal background temperature. This problem is particularly serious for very thin or membranes of nanoporous membranes with a large void fraction: these samples require higher laser powers to get a decent signal-to-noise ratio; on the other hand their smaller thermal diffusivity exacerbates the laser heating problem. One way to quantify the associated error is to estimate the temperature rise based on the absorbed laser energy [14]. However, such estimates are difficult for nanoporous membranes; besides, temperature dependence of the thermal diffusivity in nanostructures is different from that of the bulk thermal diffusivity and not known beforehand.

A more straightforward approach is to measure the effect of the pump and probe laser power on the measured thermal decay time directly. Figure 2 shows such measurements for a nanoporous Si membrane of 340 nm in thickness with a square lattice of 150 nm holes (see Sec. V) performed at background temperatures of 80 K and 298 K. One can see that the effect is measurable but moderate: a combined 4-fold increase of both pump and probe powers leads to a ~10% increase in the measured TG decay time. The measured dependencies can be extrapolated to zero laser powers to find the true value of the decay time. We chose a more conservative approach: the effect of doubling of both pump and probe power is reported as a positive error in $\tau$ or negative error in the thermal diffusivity. For example, in the case of Fig. 2, the measurement done at a pump energy of 0.26 $\mu J$ and probe power of 12 mW would yield a "laser heating" uncertainty of 6% at room temperature and 10% at 80K. These errors can be reduced by using lower laser powers at the expense of longer averaging to maintain an acceptable S/N. This fact can be observed in Fig. 3, where a signal comparison is shown between two measurements performed at different pump and probe powers. Both TTG time traces consisted of an average of 10,000 laser shots. A much larger number of averaged laser shots would be required in order to reduce the combined statistical and laser heating error due to a lower pump and probe power used in the measurements.

Another source of uncertainty, important for extremely thin membranes, is the presence of native oxide layer at the surface of Si, which is expected to have a thickness of 1 - 1.5 nm [12,31]. The presence of the native oxide does not affect the accuracy of thermal diffusivity measurements, but the measured value is the in-plane diffusivity of a multilayer structure which now includes two native oxide layers. Since the thermal conductivity of silica is much smaller than that of Si, one can assume that native oxide layers do not contribute to the thermal transport; however, they contribute to specific heat per unit area of the membrane and thus to thermal diffusivity. Consequently, the uncertainty in the native oxide thickness results in an error in the determination of the thermal conductivity of Si. This error was estimated to amount to ∼7% for a 15 nm-thick membrane but is expected to be under 1% for membrane thickness larger than 100 nm [14].

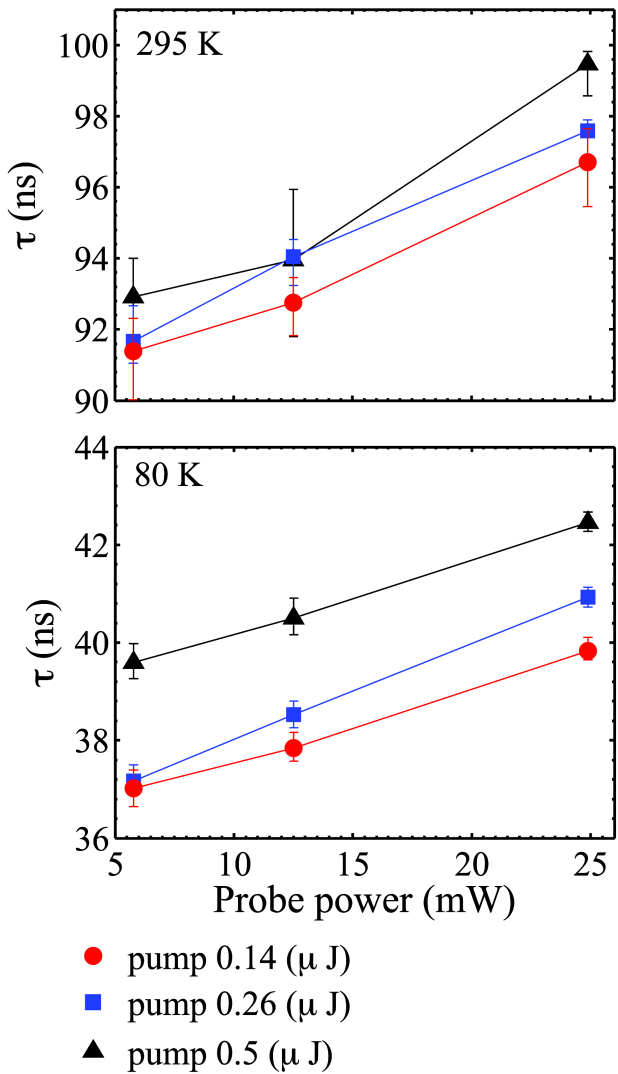

FIG. 2: Pump and probe power dependence of the measured thermal decay time at room temperature and at 80 K. Sample measured consists of a Si membrane of 340 nm thickness with a square lattice of 150 nm holes (see Sec. V). The heating effect due to the pump power becomes more prominent at cryogenic temperatures due to the low heat capacity.

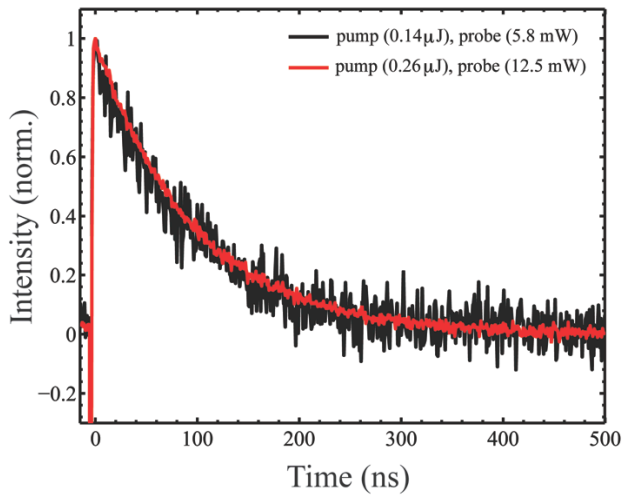

FIG 3: TTG time traces of the holey silicon membrane at different pump and probe powers. These traces were obtained with a 7.5 $\mu$m grating period.

**III NON-DIFFUSIVE TRANSPORT**

For silicon membranes with thicknesses comparable to the phonon MFP, the in-plane heat transport is normally still described by the diffusion equation, but the thermal diffusivity is reduced compared to the bulk value due to phonon scattering at the boundaries. The reduction in the thermal conductivity is described by the well-known Fuchs-Sondheimer theory [32,33] based on the Boltzmann transport equation (BTE) and initially developed for analyzing electrical conductivity of thin metal films. The TTG experiment sets another distance scale in addition to the membrane thickness, i.e., the in-plane heat transfer distance effectively equal to $L/\pi$. When this parameter becomes comparable to the phonon MFP (already reduced by the boundary scattering), in-plane transport deviates from the diffusion model, as was observed in TTG measurements on 400 nm-thick silicon membranes at room temperature [13]. Figure 4(a) shows the exponential decay rates $\tau^{-1}$ obtained from the TTG measurements versus the square of the TTG wavevector $q^2$, plotted alongside the prediction of the Fuchs-Sondheimer model for the case of diffusive in-plane thermal transport. As is readily observable, the decay rates obtained from the experiment lag below those predicted by the diffusive theory. This observed nondiffusive behavior is qualitatively consistent with the theoretical treatment of the TTG relaxation with the BTE [34,35,36]; analytical solutions were attainable due to the simplistic theoretical considerations of having a sinusoidal temperature profile [34,36]. It has been shown [34] that at relatively large TTG periods (> 1 $\mu$m) the decay dynamics should remain exponential with a decay time reduced compared to the diffusion model predictions as has indeed been observed in the experiment [13]. The exponential decay of the TTG allows one to define an "effective" thermal diffusivity $\alpha_{eff} \equiv (q^2\tau)^{-1}$, in analogy to the definition of for the diffusive case. Similarly, the "effective" thermal conductivity can be defined as $\kappa_{eff} \equiv \alpha_{eff}C_V$. Figure 4(b) shows the effective thermal conductivity (normalized to the bulk value for silicon bulk $\kappa_{bulk}$) versus TTG grating period for the two membranes studied.

The analytical theory developed in Refs. [34,35,36] could not be directly compared to these experimental data because it did not account for the boundary scattering in a membrane. More recently, a comprehensive theoretical treatment of the TTG relaxation in a thin membrane was accomplished by solving the BTE using a recently-developed deviational Monte Carlo (MC) technique [37,38,39]. The calculated results [37] are shown by solid line in Fig. 4(b). One can see that the theory still does not perfectly agree with the experimental results (however, it should be noted that the theory was entirely first-principles-based and had no fitting parameters). Further studies at smaller TTG periods, which are currently underway, will provide more material for testing theoretical models.

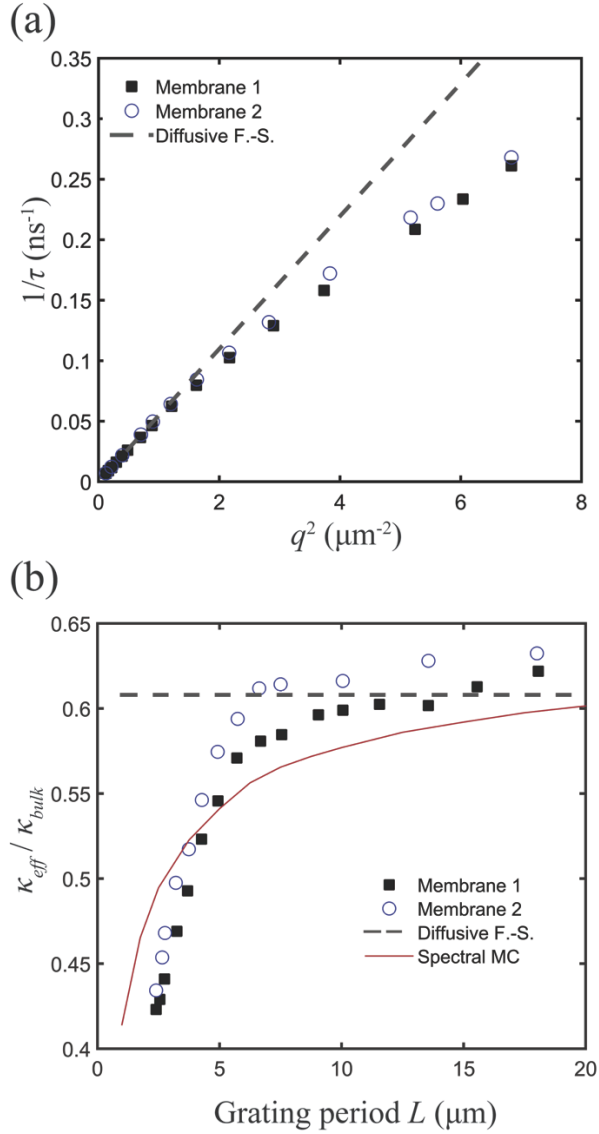

FIG. 4: (a) Decay rate versus square wavevector for two 400 nm-thick silicon membranes. The deviation from linear behavior at large wavevectors is indicative of nondiffusive thermal transport. (b) Normalized effective thermal conductivity as a function of TTG period. Spectral Monte Carlo simulation results [37] and the Fuchs-Sondheimer limit are plotted for comparison to the experimental data.

An additional insight can be provided by observing the temperature dependence of non-diffusive transport. Figure 5 shows the effective thermal diffusivity of a 200 nm-thick membrane as a function of the TTG period at temperatures from 80 to 300 K. At room temperature, transport in the 200 nm membrane at TTG periods over 7 μm is nearly perfectly diffusive, because the phonon MFP reduced by the boundary scattering is much smaller than the TTG period. Indeed, one can see that the measured diffusivity does not depend on $L$ within the range 7.5 - 21 μm. However, as the temperature is lowered, an appreciable dependence on $L$ emerges, indicating non-diffusive transport. Particularly noteworthy is the dependence at 80 K, where $\alpha(L)$ does not level off to a constant value even at 21 $\mu$m. It is generally expected that phonon transport should become non-diffusive at low temperatures as the phonon MFP increases. However, in a thin membrane the phonon MFP is additionally suppressed by the boundary scattering. As will be shown in the next section, room temperature thermal transport in thin Si membranes is consistent with the assumption of diffuse (zero specularity) boundaries. At low temperatures the wavelength of heat-carrying thermal phonons increases and surfaces become specular as is well known from experiments conducted at liquid He temperatures [40]. The analysis of the temperature-dependent TTG data is still underway but the fact that non-diffusive transport is observed at a TG period more than hundred times larger than the membrane thickness may indicate an increase in the surface specularity already at 80 K.

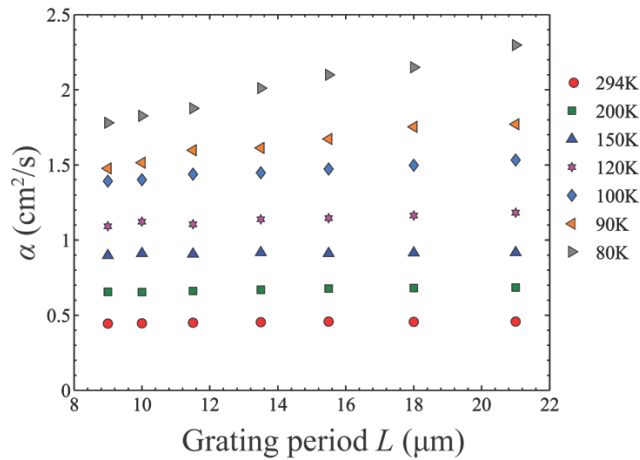

FIG. 5: Measured effective thermal diffusivity values for a 200 nm silicon membrane as a function of the TTG period for temperatures from 80K to room temperature. A dependence on $L$ emerges at low temperatures.

**IV. THICKNESS DEPENDENCE OF THE THERMAL CONDUCTIVITY OF SILICON MEMBRANES**

As explained in the preceding section, at long TG periods room temperature in thin Si membranes is diffusive, but the thermal diffusivity is reduced compared to its bulk value due to boundary scattering. The in-plane thermal conductivity in free-standing Si films was measured as a function of the film thickness from 15 nm to 1518 nm with TTG periods ranging from 11 to 21 $\mu$m at 294K [14]. The measured thickness dependence of the conductivity is shown in Fig. 6 alongside the literature data. It should be noted that most previous measurements (with the exception of Refs. [9,13]) were done on supported films rather than on free-standing membranes. Furthermore, many previous measurements had large error bars; in fact for measurements done on membranes with thickness over 1 $\mu$m the error bars were such that it was not possible to conclude whether the measured conductivity was bulk value. The TTG measurements significantly reduced the error bars, with the size effect now clearly visible all the way to ~1.5 $\mu$m. Furthermore, it provided a set of measurements spanning two orders of magnitude of the membrane thickness. These advances made it possible to use the TTG data to do test first-principle-based calculations of thermal conductivity in a nanostructure. Shown by a solid line in Fig. 6 are calculations done with the Fuchs-Sondheimer model (assuming diffuse boundaries) coupled with first-principles calculations of phonons MFPs [41]. The theoretical calculations are in very good agreement with the TTG measurements. The agreement indicates that the Fuchs-Sondheimer model, which assumes bulk-like phonon propagation in the body of the structure, is valid in real nanostructured membranes as thin as 15 nm.

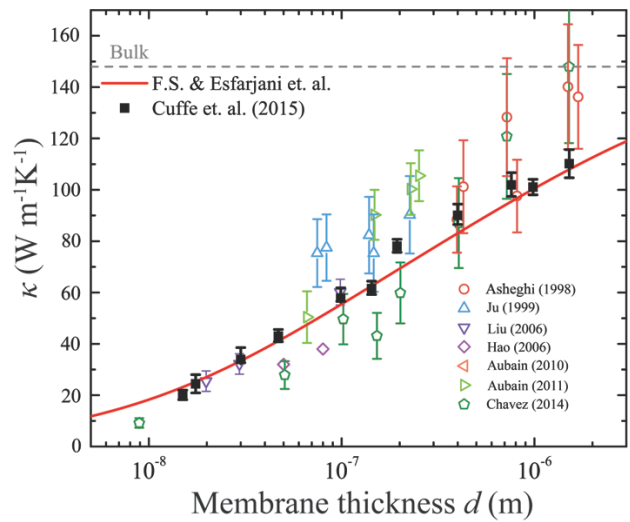

FIG. 6: Measured thermal conductivity $\kappa$ as a function of membrane thickness $d$. Black solid squares correspond to the experimental points from Cuffe et al. [14]. Solid red line corresponds to theoretical calculations employing the Fuchs-Sondheimer model and first-principles-calculated phonon lifetimes. Dashed horizontal line corresponds to the bulk thermal conductivity value for Si. Other measurements of thermal conductivity for supported and unsupported Si thin films are presented in open symbols. [8,9,42,43,44,45,46].

TTG measurements shown in Fig. 6 can also be used to solve the inverse problem of reconstructing the phonon MFP distribution (i.e. the thermal conductivity accumulation function vs MFP) [47]. Such reconstruction has been demonstrated [14] and showed a good agreement with the MFP distribution calculated from first-principles [41]. It should be noted that in a number of recent studies the reconstruction of the MFP distribution was pursued based on measurements of non-diffusive transport as a function of the heat transfer distance [47,48,50] essentially analogous to measurements described in Sec. III, but requiring nanoscale metal heaters [49,50] in order to extend the measurements to submicron distances. The advantage of the approach used in Ref. [14] based on measuring *diffusive* transport as a function of a nanostructure dimension was that (i) it relied on a rigorous solution to the direct problem provided by the Fuchs-Sondheimer model and (ii) measurements covered a wide range (two orders of magnitude) of the length parameter.

**V. HOLEY SILICON MEMBRANES**

Nanostructuring has been identified as an effective way to inhibit phonon-mediated thermal transport, thereby enhancing thermoelectric performance with respect to the native unstructured material. Due to the prospect of using this design strategy to enhance the otherwise poor prospects of crystalline silicon-based devices for thermoelectric applications, there has been much recent fundamental interest in the nature of thermal transport in nanoporous, or "holey" silicon (HS) membrane structures [17]. While for solid Si membranes thicker than 10 nm the reduction in thermal conductivity compared to bulk Si does not exceed an order of magnitude (see Fig. 6) holey silicon membranes with the critical dimension on the order of 20 nm yielded up to two orders of magnitude reduction in thermal conductivity at room temperature [6]. However, the origin of this reduction is still debated. While several studies suggested the observations could not be explained without invoking phononic bandgaps arising in the periodic nanopore structure [6,7,51], other studies indicated that the phononic effects in HS at room temperature are unlikely [52,53]. Furthermore, there are significant discrepancies between measurements done on structures with similar dimensions [17]. These unresolved issues underscore the need for measurements that employs a technique with inherently high absolute accuracy such as TTG.

We used the TTG technique to investigate room temperature thermal transport properties of a HS structure fabricated in a 340 nm-thick membrane. Fig 7 (a) shows SEM image from the sample. It has 135 nm hole diameter, 206 nm periodicity and a limiting dimension (neck size) of 70 nm. Measurements in the range of TTG periods 3.2-11.5 $\mu$m presented in Fig. 7(b) showed that the TTG decay rate $1/\tau$ is proportional to $q^2$ both at room temperature and at 84 K, in perfect agreement with Eq. (2). This linear scaling indicates diffusive transport in contrast to solid membranes of similar thickness where significant deviations from the linear scaling were observed (compare with Fig. 4(a)). The fact that the perfectly diffusive behavior is observed even at 84K suggests a strong reduction of the phonon MFP due to diffuse scattering by the pores.

The measured room temperature in-plane thermal diffusivity was $0.15\ cm^2 s^{-1}$ and the thermal conductivity $16.5\ Wm^{-1}K^{-1}$, almost an order of magnitude smaller compared to the bulk value and about 5 times smaller compared to a solid membrane of a similar thickness. The thermal conductivity was calculated using the known relation $\kappa = C\alpha(1-\phi)$ where $C = 1.64x10^6 Jm^{-3}K^{-1}$ is the volumetric heat capacity [54], $\alpha$ is the measured thermal diffusivity and $\phi = 0.337$ is the porosity of the sample.

To compare the measurement with theory, we calculated the thermal diffusivity using a "kinetic-type" Monte Carlo method [55,56] based on the linearization of the deviational Boltzmann transport equation. As mentioned in [55], the study of a periodic nanostructure with this Monte Carlo approach introduces a deterministic and a stochastic uncertainty, both of which are controllable. The deterministic uncertainty originates from the truncation of particle trajectories which otherwise would be infinite, and is estimated to be lower than 2%. We used $10^8$ computational particles, with a resulting statistical uncertainty below 0.1%. The inputs, namely the dispersion relation and frequency-dependent relaxation times, were obtained from ab-initio calculations [41]. In order to account for isotope scattering, we added to the ab-initio relaxation rates a term of the form $Ag\omega^2$, where $g$ is the density of states and $A$ was calculated following the procedure outlined in [57] and assigned the value $A = 1.187e^{-32}\ m^3$. The obtained thermal conductivity was $19.7\ Wm^{-1}K^{-1}$, that corresponds to a thermal diffusivity of $0.18\ cm^2 s^{-1}$, agreeing reasonably well with the measured thermal diffusivity values. Fig. 7 (c) shows a comparison between the measured TTG signal and one-dimensional thermal exponential decay using $\alpha$ obtained from the MC simulations.

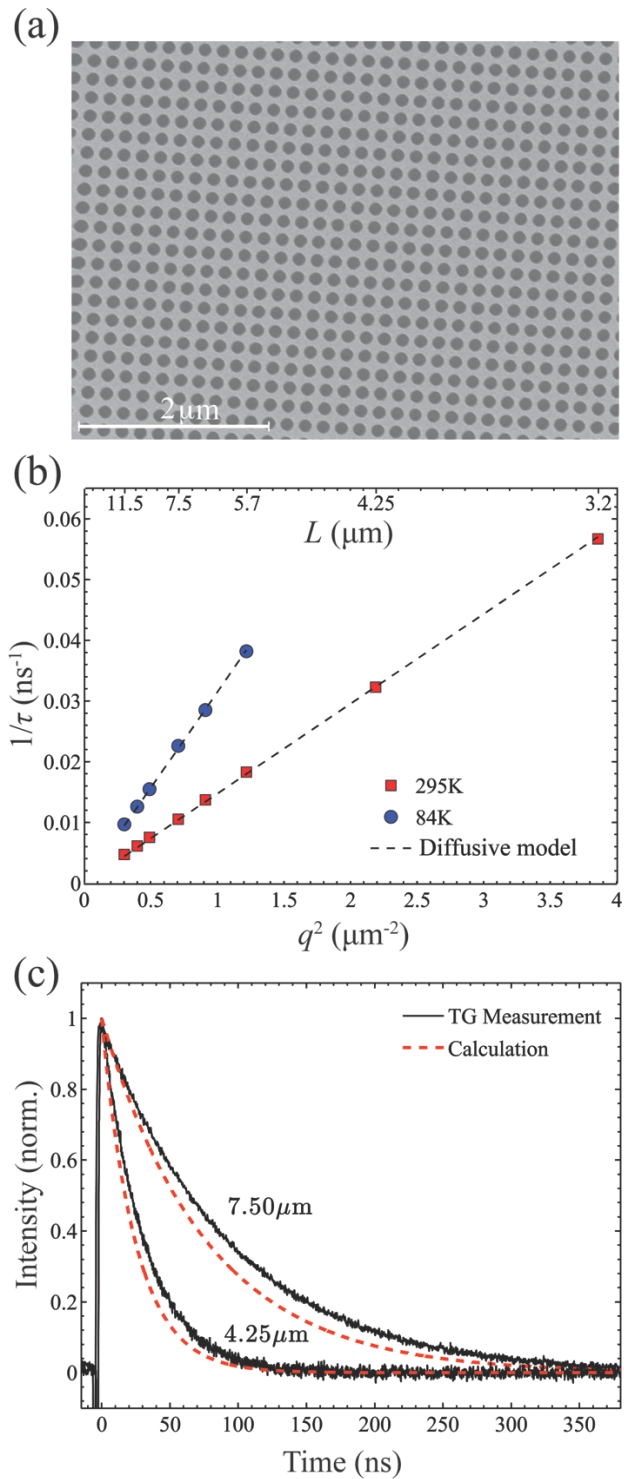

Fig. 7 (a) SEM image of the holey Si sample (135 nm hole diameter, 206 nm periodicity, 340 nm thickness). (b) Thermal grating decay rate versus the wavevector magnitude squared showing diffusive behavior at 295 K and 84 K. (c) Comparison between TTG signal and the calculated exponential decay using the obtained thermal diffusivity from MC simulations.

**VI. CONCLUSIONS AND OUTLOOK**

We have seen that the application of the TTG method to studying thermal transport in silicon membranes already resulted in some important findings. (i) It permitted a direct observation of nondiffusive phonon-mediated transport at room temperature. "Direct" here means that the deviation from the diffusion model could be seen directly from the deviation of the measured thermal decay rate from a quadratic dependence on the TTG wavevector, without any further analysis. It should be noted, that this experiment could be as well done on a bulk sample [58]; thin membranes were used for technical reasons, to ensure one-dimensional thermal transport to simplify the interpretation of the measurements. (ii) Measurements of thickness-dependent thermal conductivity of suspended membranes across a wide thickness range showed a good agreement with the Fuchs-Sondheimer model employing first-principles-calculated phonon lifetimes and assuming diffuse scattering at the boundaries. The introduction of first-principles calculations of the lattice thermal conductivity based on density functional theory potentials [59] was an important milestone in studying thermal transport. However, the comparison with experiment has been limited to bulk materials [59,60]. Now we have demonstrated that first-principles-based theory with no fitting parameters is capable of predicting the size effect in Si nanostructures. (iii) In contrast to solid membranes, a nanoporous membrane yielded fully diffusive transport indicating diffuse scattering by the pores. The room temperature thermal diffusivity of a holey membrane reasonably well agrees with calculations based on BTE with diffuse boundaries.

Although the data from holey Si are still limited and a study on a systematic set of sample with varying geometry of nanopores is desirable, the results obtained to date indicate that room temperature thermal conductivity of both solid and holey membranes is well described by the BTE with diffuse boundaries, i.e. by the model going back to Casimir [61] and Fuchs [32]. This conclusion is likely to be valid for other Si nanostructures with critical dimension down to at least 15 nm. We have not seen any phonon interference effects such as phononic bandgaps in a periodic lattice of nanopores or confined phonon modes in a membrane [62]. Indeed, phonon interference effects arise due to specular reflections at the boundaries, while our results indicate that the boundaries are effectively diffuse. Fabricating a nanostructure with atomically perfect specular boundaries remains a challenge for further research. Furthermore, boundaries will become specular at low temperatures when the wavelength of thermal phonons gets much larger than the height of surface roughness. At temperatures below 4K, phonon interference and confinement definitely influence thermal transport [10]. The question at what temperatures the surface specularity becomes significantly non-zero for heat-carrying thermal phonons remains open. While data shown in Fig. 5 may indicate an effect of non-zero specularity at 80K, further analysis coupled with further temperature-dependent measurements on membranes with different thicknesses is needed to address this question.

Further progress in studying nanoscale thermal transport is being made with a variety of techniques including both device-based approaches and optical methods such as TDTR [21] and Raman thermometry [63] as well as TTG. While each technique has its advantages and challenges, the results described above indicate that TTG is particularly well suited for measurements on 2D nanostructures such as thin membranes. Extending its reach to other 2D structures and 2D materials would be a logical next step. Exciting new avenues are also being opened by coherent x-ray and extreme ultraviolet sources, which both open new ways for studying phonons [64] and thermal transport and create new opportunities for existing techniques such as TTG with nanoscale spatial periods [65].

## VII. ACKNOWLEDGMENTS

The work done at MIT was supported as part of the "Solid State Solar-Thermal Energy Conversion Center (S3TEC)," an Energy Frontier Research Center funded by the U.S. Department of Energy, Office of Science, Office of Basic Energy Sciences under Award No. DE-SC0001299/DE-FG02-09ER46577. The contribution by A.V.-F. and J. J. A.-G. was partially supported by Project 192 "Fronteras de la ciencia" and Project 251882 "Investigación Científica Básica ". A. V.-F. also appreciates support from Conacyt through normal and mixed scholarships. MS and CMST acknowledge support from the Spanish program Severo Ochoa (Grant SEV-2013-0295), projects PHENTOM (FIS2015-70862-P) and nanoTHERM (CSD2010-00044), as well as from the EU project MERGING (309150). Z.L. and E.N.W. further acknowledge support and funding from the Air Force Office of Scientific Research (AFOSR), and are grateful to program manager Dr. Ali Sayir.